# Distributed Data and Programs Slicing

Mohamed A. El-Zawawy[1,2]

[1]College of Computer and Information Sciences, Al Imam Mohammad Ibn Saud Islamic University (IMSIU),
Riyadh, Kingdom of Saudi Arabia
[2]Department of Mathematics, Faculty of Science, Cairo University, Giza 12613, Egypt
maelzawawy@cu.edu.eg

**Abstract:** This paper presents a new technique for data slicing of distributed programs running on a hierarchy of machines. Data slicing can be realized as a program transformation that partitions heaps of machines in a hierarchy into independent regions. Inside each region of each machine, pointers preserve the original pointer structures in the original heap hierarchy. Each heap component of the base type (e.g., the integer type) goes only to a region of one of the heaps. The proposed technique has the shape of a system of inference rules. In addition, this paper presents a simply structure type system to decide type soundness of distributed programs. Using this type system, a mathematical proof that the proposed slicing technique preserves typing properties is outlined in this paper as well.
[El-Zawawy MA. **Distributed Data and Programs Slicing**. *Life Sci J* 2013;10(4):1361-1369]. (ISSN: 1097-8135).
http://www.lifesciencesite.com. 180

## 1. Introduction

Breaking down a large distributed program into smaller pieces or minimizing its size is essential for many software analysis techniques such as parallelization [19], debugging [22], program comprehension [14], testing [16], downsizing, and restructuring. Introduced by Mark Weiser [20], data and program slicing [3] are applicable techniques for narrowing the focus of a program to a certain region of the memory. Interesting enough, data slicing was motivated by the desire of guiding students through debugging programs. A program slice can be defined as an executable subset of the program that simulates the original program on a certain data slice (region of the memory). Data slicing is useful when compilers need to modify data structures in the program being compiled without breaching pre-compiled assumptions about data layout.

Distributed systems [15] are typically constructed on hierarchical memory structure (Local stores and caches of processors, such as cell game processor, are organized in a hierarchal fashion). This memory system equips each process with explicitly controlled local caches or stores. Distributed computations and their hierarchy models of memories have been the focus of much research activities. This is partially justified by the existence of low-cost processors facilitating building such distributed systems. Rather than distributed systems of a number of processors create immunity against different sorts of failures, they also have incremental growth capabilities and high throughput. One example of memory hierarchy is to partition memory into computational grids consisting of clusters partitioned into nodes including program threads. Practically, memories of most distributed programming languages have a two level abstraction. The model of distributed systems used in this paper is the *single program multi data (SPMD)* model [15].

Type systems [13] are theoretical means for proving type soundness which amounts to the absence of method-not-found and field-not-found errors. For distributed programming languages, like $\mathcal{DLang}$ of this paper, a type system guarantees that every use of a location considers its predefined type. Proving this property is, however, not easy due to the potential intervention of executing the program on different machines of hierarchy.

This paper presents a new technique for slicing distributed programs running on hierarchal memories. The proposed technique has the form of inference rules which are simply structured. The new technique is illustrated using a simple, however rich, model of distributed programming language ($\mathcal{DLang}$ - Figure 2). The paper also presents a type system that checks type soundness of programs of $\mathcal{DLang}$ and programs resulting from the proposed slicing technique. A prove that prosperity of type soundness of a program is preserved by the slicing technique is also presented in this paper.

Rather than the traditional algorithmic way, using a system of inference rules [4–6] to achieve static analyses and transformations of distributed programs has recently proved to be a good choice. This is so as derivations in the system of rules work as simply-structured correctness proofs for results of the system. Such proofs are required by many applications like proof-carrying code [9].

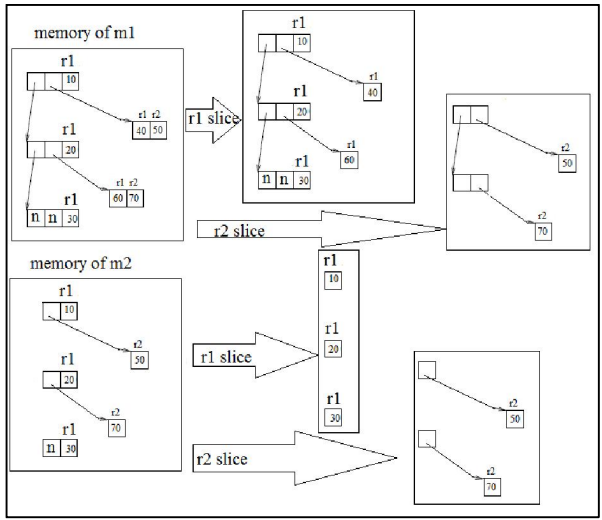

**Fig. 1**. A motivating example for distributed data and program slicing.

**Motivation**

Suppose a distributed system including two machines $m_1$ and $m_2$ each of them has two memory regions $r_1$ and $r_2$. On this system, suppose also a distributed program defining the following types:

$type\ t_1 = struct$
$\{y_1: ptr_1^1, y_2: ptr_2^1, y_3: ptr_2^2, y_4: int(r_1, \{m_1, m_2\})\};$
$type\ t_2 = struct$
$\{y_1: int(r_1, \{m_1\}), y_2: int(r_2, \{m_1, m_2\})\};$

The left-hand-side of Figure 1 illustrates how these data types will be allocated on memories of machines $m_1$ and $m_2$. The right-hand-side of the figure explains distributed data slicing effects on the two machines into the two regions. Annotations are used in integer fields of data structures to determine regions of machines that will host these fields. The slicing includes inserting shapes of original data structures in each region of each machine. However such a shape of a region typically contains only relevant data and necessary pointers.

The aim of this paper is to provide formal techniques to transform memories of $m_1$ and $m_2$ into that on the right-hand-side Figure 1. Moreover the proposed technique is required to slice the program that includes the type definitions into slices each works on the data of a certain region.

**Contributions**

Contributions of this paper include:

1. A new type system for checking type soundness of distributed programs.
2. A novel and sound technique for slinging distributed data and programs running on hierarchal systems of memories.

3. A mathematical proof that the proposed slicing technique preserves typing properties.

**Organization**

The organization in the rest of the paper is as following. The syntax of the target language and a system of inference rules for type checking of the language constructs are presented in Section 2. Section 3 introduces the new technique for slicing of distributed programs running on hierarchal machines. The type system of Section 2 is used in Section 3 to prove that the proposed slicing technique preserves type-soundness of sliced programs. Related work is reviewed in Section 4.

$x, y \in lVar, n \in \mathbb{Z}, i_{op} \in \mathbb{I}_{op}, and\ b_{op} \in \mathbb{B}_{op},$
$\quad and\ m \in M \subseteq \mathcal{M}$
$\tau \in Types ::= int(r_i, M) \mid ptr^m \tau \mid t$
$\qquad\qquad\qquad \mid struct\{y_1: \tau_1, \ldots, y_n: \tau_n\}.$
$d \in Defs ::= type\ t = t \mid t_1 t_2 \mid \varepsilon.$
$l \in lexpr ::= x \mid l. y \mid * e.$
$e \in Expr ::= l \mid e_1\ i_{op}\ e_2 \mid \& l \mid new\ \tau \mid$
$modify - w(e, m) \mid compute\ e\ at\ m \mid cast\ <$
$ptr^m \tau \to int(r_i, M) > e \mid cast\ < int(r_j, M_j) \to$
$int(r_i, M_i) > e.$
$S \in Stmts ::= skip \mid l := e \mid\ compute\ S\ at\ m$
$\qquad\qquad\qquad \mid\ S_1; S_2 \mid if\ e\ then\ S_t else\ S_f$
$\qquad\qquad\qquad \mid while\ e\ do\ S_t.$
$p \in Progs ::= dS$

**Fig. 2**. The programming language model, $\mathcal{DL}ang$.

## 2. Target Language $\mathcal{DL}ang$: Syntax and Type Checking

This section presents the syntax of the target language and a system of inference rules for type checking of the language constructs. The syntax of the target programming language, $\mathcal{DL}ang$, is presented in Figure 2. We assume an arbitrary hierarchy of machines. In term of parallel programs, a single execution thread resembles a machine. A countably infinite collection of variables (machine-private) is used in the language and denoted by $lVar$ with typical elements $x, y$. Sets of finite arithmetic and Boolean operations are denoted by $\mathbb{I}_{op}$ and $\mathbb{B}_{op}$, respectively. The set of machine identifiers is denoted by $\mathcal{M}$. The types of the language are integer, pointer, structure, and named types ($t$). The empty structure, $struct\{\}$, is denoted by $void$ as a shorthand. It is noted that base types (integer types) are quantified (Similar type quantification can be found in [3]) with pairs of a region and a set of machines.

This is so to determine the locations of the data of these types. The language programs are executed on a distributed system of $\delta$ machines with *identifiers* $m_1$ through $m_\delta$. Each machine has $\alpha$ regions named $r_1$ throught $r_\alpha$. The set of all the regions is denoted by $\mathfrak{R}$. $\mathcal{DLang}$ can be realized as a generalization of the language in [10].

The hierarchy level hosting the smallest common ancestor of two machines is the *distance* between the two machines. The number of levels in the machines hierarchy is dubbed the *depth* of the hierarchy. The *width* of a pointer on a machine is the distance between the machine hosting the pointer and the machine hosting the location pointed-at by the pointer. We assume a function *width-f* that assigns each pointer its width. The symbol $h$ denotes the height of the machine hierarchy. Therefore the set of widths of pointers is $\{1, \dots, h\}$ and the pointer type is parameterized by the machine *id* of the location it points-at.

Expressions ($e$) and l-expressions ($l$) are inspired by that of $C$. Binary operations (arithmetic and boolean) are only applicable to integers in the same region of possibly different machines. Expressions include:

- *new $\tau$*: allocates a memory location of type $\tau$ and returns the allocated address.
- *modify $-w(e, m)$* : changes the pointer width. More specifically, it changes the machine the expression points-at. Hence the type of $e$ becomes the reference $ptr^m\tau$, rather than $ptr^{m'}\tau$.

- *compute $e$ at $m$*: computes the expression $e$ on the machine $m$ and distributes the value to other machines.
- *cast $< ptr^m\tau \rightarrow int(r_i, M) > e$:* casts from pointers to integers in different regions of different machines.
- *cast $< int(r_j, M_j) \rightarrow int(r_i, M_i) > e$*: casts between integers in different regions of different machines.

The statement *compute $S$ at $m$* executes the statement $S$ on the machine $m$. A $\mathcal{DLang}$ program is a sequence of type definitions followed by a statement.

**Remark 1**. Primitive values such as Boolean values and integers are not incorporated in the language $\mathcal{DLang}$ . It is straightforward to include them as an extension. Although the language $\mathcal{DLang}$ is *SPMD*, the proposed data-slicing technique is easily extendable to other parallelism models. However these extensions are not considered in this paper.

A system of inference rules for type checking of components of $\mathcal{DLang}$ is presented in Figures 3 and 4. A *context*, $\Gamma$, is a map from variables and type names to types. A program is well-typed (WT) if its body, $S$ , is well-typed with initial environment $\Gamma_f$. In the allocation rule $(new_t)$, the expression generates a pointer type $ptr^m$ for all $m$. As the allocation takes place on the same machine that is executing the statement, the width of the created pointer is 1.

$$\frac{\Gamma \vDash_l \; l : \tau}{\Gamma \vDash_e \; l : \tau}(l^t) \qquad \frac{\Gamma \vDash_e \; e_1 : int\,(r_i, M) \qquad \Gamma \vDash_e \; e_2 : int\,(r_i, M)}{\Gamma \vDash_e \; e_1 \, i_{op} \, e_2 : int\,(r_i, M)}\left((e_1 \; i_{op} \; e_2)^t\right)$$

$$\frac{m \in \mathcal{M}}{\Gamma \vDash_e \; new \; \tau : ptr^m \; \tau}(new^t) \qquad \frac{\Gamma \vDash_l \; l : \tau}{\Gamma \vDash_e \; l : ptr^m \; \tau}(\&l^t) \qquad \frac{\Gamma \vDash_e \; e : \tau' \qquad \tau' \subseteq \tau}{\Gamma \vDash_e \; e : \tau}(\subseteq^t)$$

$$\frac{width - f(e) = m' \qquad \Gamma \vDash_e \; e : ptr^{m'} \; \tau}{\Gamma \vDash_e : modify - w\,(e, m) \; : \; ptr^m \; \tau}(modify - w^t) \qquad \frac{\Gamma \vDash_e \; e : \tau}{\Gamma \vDash_e \; compute \; e \; at \; m : \tau}(comp^t)$$

$$\frac{\Gamma \vDash_e \; e : int(r_j, M_j)}{\Gamma \vDash_e \; cast \; < int(r_j, M_j) \hookrightarrow int(r_i, M_i) > e : int(r_i, M_i)}(cast_1^t)$$

$$\frac{\Gamma \vDash_e \; e : ptr^m \, \tau(r_i, M) \; has \; a \; type \; reachable \; form \; \tau}{\Gamma \vDash_e \; cast < ptr^n \tau \hookrightarrow int(r_i, M) > e : int(r_i, M)}(cast_2^t)$$

**Fig. 3.** Typing rules for expression.

The *modify* expression enables modifying the machine *id* that a pointer references. This, of course, results in changing the top-level expression width. This way is used to decrease the width of the expression more often than to increase it. This is so because of the subtyping rule. After a dynamic analysis, the *modify* statement can be used to express that the expression references data on a machine closer than initially thought. The inclusion relationship in the rule ($\subseteq_t$) applies only on *structure* types. The remaining rules are self-explanatory.

Hence if, for example, the data of a liked list in the original heap hierarchy are annotated with different regions of different machines, then every region on each machine will include a similar linked list with only the list data for this region. Hence whereas base fields will be divided among regions of different machines in the hierarchy according to their annotations, the same pointer maybe replicated into many regions as necessary. Therefore in data slicing base fields are separated and pointers are replicated.

$$\overline{\Gamma \vDash_t \epsilon : WT} \; (\epsilon^t) \qquad \frac{\Gamma(t) = \tau}{\Gamma \vDash_t \; \text{type } t = \tau : WT} \; (t^t) \qquad \frac{\Gamma \vDash_t \; d_1 : WT \qquad \Gamma \vDash_t \; d_2 : WT}{\Gamma \vDash_t \; d_1 \; d_2 : WT} \; (t^t)$$

$$\frac{x \in dom(\Gamma)}{\Gamma \vDash_l x : \Gamma(x)} (x_1^t) \qquad \frac{x \in dom(\Gamma)}{\Gamma \vDash_l x . (r_i, m) : \Gamma(x)} (x_2^t) \qquad \frac{\Gamma \vDash_l \; y : \tau_1 \qquad \{y : \tau_2\} \subseteq \tau_1}{\Gamma \vDash_l \; l . y : \tau_2} (l . y^t)$$

$$\frac{\Gamma \vDash_e \; e : ptr^m \tau}{\Gamma \vDash_l * e : \tau} (* e^t) \qquad \overline{\Gamma \vDash_s \; skip : WT} \; (skip^t) \qquad \frac{\Gamma \vDash_l \; l : \tau \qquad \Gamma \vDash_e \; e : \tau}{\Gamma \vDash_s \; l := e : WT} (:=^t)$$

$$\frac{\Gamma \vDash_s \; S : WT}{\Gamma \vDash_s \; \text{compute } S \text{ at } n : WT} (compute^t) \qquad \frac{\Gamma \vDash_s \; S_1 : WT \qquad \Gamma \vDash_s \; S_2 : WT}{\Gamma \vDash_s \; S_1 S_2 : \; WT} (Seq^t)$$

$$\frac{\Gamma \vDash_e \; e : int(r_i, M) \qquad \Gamma \vDash_s \; S_t : WT \qquad \Gamma \vDash_s \; S_f : WT}{\Gamma \vDash_s \; \text{if } e \text{ then } S_t \text{ else } S_f : WT} (if^t)$$

$$\frac{\Gamma \vDash_e \; e : int(r_i, M) \qquad \Gamma \vDash_s \; S_t : WT}{\Gamma \vDash_s \; \text{while } e \text{ do } S_t : \; WT} (wle^t)$$

**Fig. 4.** Typing rules for type definitions, left expressions, statements, and programs.

### 3. Data Slicing of $\mathcal{DLang}$

This section presents a new technique for data slicing [3] of distributed programs [10] running on hierarchal machines. The type system of the previous section is used later in this section to prove that the proposed technique preserves type-soundness of sliced programs. Data slicing of distributed programs aims at dividing heaps of hierarchy (distributed) machines into separate regions. The base types (only integers in our case) of the input program of data slicing have to be region-machine-annotated. The result of the slicing is a new program whose data structures are split into separate regions of the heaps of hierarchy machines. Of course the new and original programs have to be semantically equivalent. Data must be contained in regions of machine hierarchy according to data annotations; adding new pointers to do so is allowed. However complete independence of regions is assumed; cross-region boundaries pointers are not allowed.

As a result of data slicing, every machine region reflects the original structure of heaps in machine hierarchy (for an example see Figure 1).

**Example:** Consider applying the slicing technique on the statement compute:

$$\text{compute } * x . y_1 \text{ at } m_1,$$

where the type of $x$ is $ptr^1 t_2$ and $t_2$ is the type defined in the motivating example illustrated by Figure 1. The result of the slicing will be the statement:

$$\text{compute } * x . (r_1, m_1) . y_1 \text{ at } m_1.$$

This amounts to returning the value of $y_1$ of structure $x$ hosted by region $r_1$ of machine $m_1$.

A simple lemma induction, on structure of type's $\tau$, proves Lemma 1 reasoning about type transformations of Figure 5.

**Lemma 1.** *Suppose that* $\tau \rightsquigarrow_{(m,i)}^{\tau} \tau'$. *Then* ($\tau' \neq void$) *implies* $\tau$ *and* $\tau'$ *are of the same type.*

Using Lemma1, it is not hard to prove Corollary 1, describing transformations of Figure 6.

**Corollary 1.** *Suppose that* $d \rightsquigarrow_{(m,i)}^{\tau} d'$. *Then* $d$ *and* $d'$ *define equivalent types.*

Figures 5, 6, and 7 present the proposed slicing technique. Inference rules for slicing types over regions and machines of a distributed system are shown in Figure 5. The main notation in this figure is $\tau \leadsto^{\tau}_{(m,i)} \tau'$ meaning that the type $\tau'$ is the slice of the original type $\tau$ on the region $r_i$ of the machine $m$. The rules $(int_1^s)$ and $(int_2^s)$ express that the slice of region $r_i$ of machine $m$ includes only integers annotated with the pair $(m, i)$. The rules $(ptr_1^s)$, $(ptr_2^s)$, and $(str^s)$ for pointers, and structures recursively invoke the inference rules of the figure. For a machine $m$, the rule $(str^m)$ calculates the slices of a type $\tau$ on the regions of the machine.

the rule only involves moving a single integer between regions. On the other side, as clarified by the rule $(cast_2^s)$, the transformation of an expression of pointer casting is affected by the casting being region-specific. Rules for slicing over machines of left expression, statement, and program are presented in figure 8. According to the rule $(l^s)$ slicing a reference to a variable amounts to selecting the element of the variable belonging to the addressed region. A key rule in the proposed technique is that of

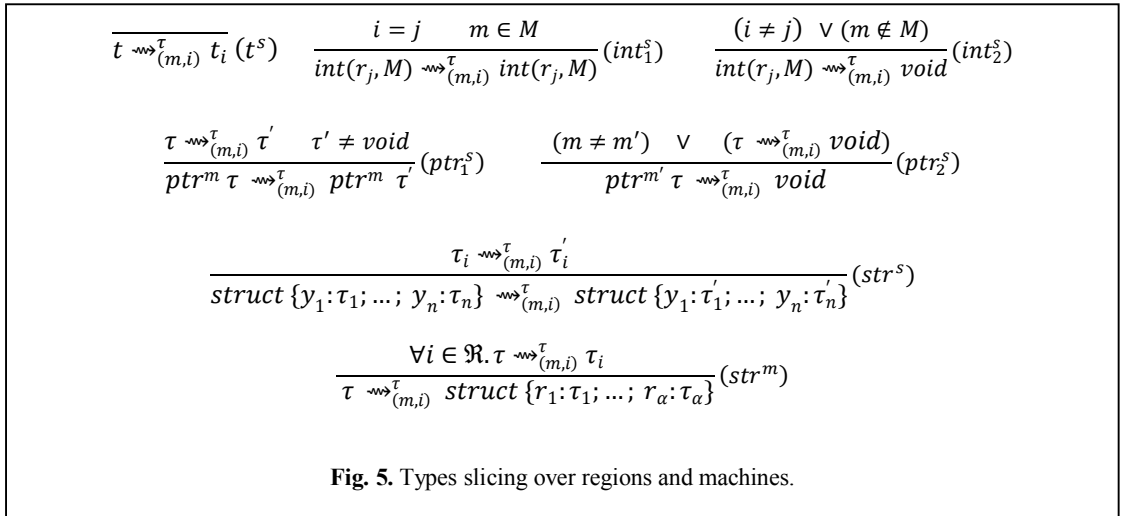

**Fig. 5.** Types slicing over regions and machines.

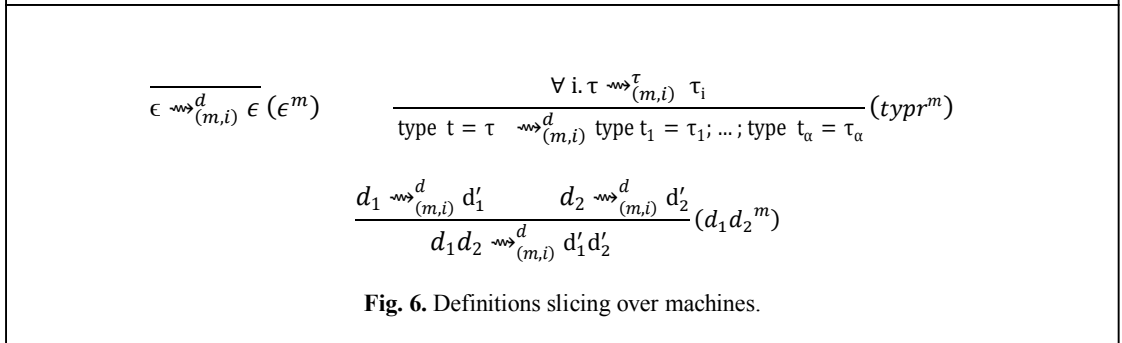

**Fig. 6.** Definitions slicing over machines.

Figure 6 shows rules for slicing definitions over regions of machines. The rule $(typr^m)$ is the basic one for definition slicing. In this rule, the right-hand-side of a type definition is sliced over different regions of a machine $m$ transforming the original definition statement into $\alpha$ statements on $m$.

Inference rules for expression slicing over regions are included in Figure 7. The rule $(cast_1^s)$ recursively uses expression rules to compute an integer in region $r_j$. This integer is then casted into this region. Clearly the assumption that inter-region pointers are not allowed is preserved by this rule as assignment $(:=^s)$. The main idea behind this rule is to achieve, if the assigned type is not void, the corresponding assignment in every region of every machine.

**Remark 2.** The fact that assignments are done separately in different regions of different machines is the reason that most of the inference rules on the proposed technique rules are region-oriented. This is done assuming the existence of the expression sliced-type in the addressed region.

The following results prove that the proposed data slicing technique preserves type-checking properties of distributed programs.

$$\frac{l \rightsquigarrow^l_{(m,i)} l'}{l \rightsquigarrow^e_{(m,i)} l'}(l^s) \qquad \frac{e_1 \rightsquigarrow^e_{(m,i)} e'_1 \quad e_2 \rightsquigarrow^e_{(m,i)} e'_2}{e_1 \; i_{op} \; e_2 \rightsquigarrow^e_{(m,i)} e'_1 \; i_{op} \; e'_2}((e_1 i_{op} e_2)^s)$$

$$\frac{\tau \rightsquigarrow^\tau_{(m,i)} \tau'}{new \; \tau \rightsquigarrow^\tau_{(m,i)} new \; \tau'}(new^s) \qquad \frac{l \rightsquigarrow^l_{(m,i)} l'}{\& \; l \rightsquigarrow^e_{(m,i)} \& \; l'}(\& l^s)$$

$$\frac{e \rightsquigarrow^e_{(m,i)} e'}{modify-w(e,d) \rightsquigarrow^e_{(m,i)} modify-w(e',d)}(modify-w^s)$$

$$\frac{e \rightsquigarrow^e_{(m,i)} e'}{compute \; e \; at \; n \rightsquigarrow^e_{(m,i)} compute \; e' \; at \; n}(comp^s)$$

$$\frac{e \rightsquigarrow^e_{(m,i)} e'}{cast < int(r_j, M_j) \to int(r_i, M_i) > e \rightsquigarrow^e_{(m,i)} cast < int(r_j, M_j) \to int(r_i, M_i) > e'}(cast^s_1)$$

$$\frac{\tau \rightsquigarrow^\tau_{(m,i)} \tau' \quad e \rightsquigarrow^e_{(m,i)} e'}{cast < ptr^m \tau \to int(r_i, M) > e \rightsquigarrow^e_{(m,i)} cast < ptr^m \tau' \to int(r_i, M) > e'}(cast^s_2)$$

**Fig. 7.** Expression slicing over regions.

**Lemma 2.** *Suppose that* $l \rightsquigarrow^l_{(m,i)} l'$ *and* $\Gamma \vDash_l l{:}\tau$. *Then* $\Gamma \vDash_l l'{:}\tau$.

**Proof.** The proof is by structure induction on the structure of left expressions, $l$, as follows.

- The case $l = x$: in this case $l' = x.(r_i, m)$. By rules $(x^t_1)$ and $(x^t_1)$, $\Gamma \vDash_l l{:}\Gamma(l)$ and $\Gamma \vDash_l l'{:}\Gamma(l)$. Hence the required is satisfied.

- The case $l = l.y$: in this case $l' = l'.y$. By the rule $(l^s)$, it is true that $\rightsquigarrow^l_{(m,i)} l'$. Now since $\Gamma \vDash_l l.y{:}\tau$, then by rule $(l.y^t)$ there exists $\tau_1$ such that $\Gamma \vDash_l l{:}\tau_1$ and $\{y{:}\tau_1\} \subseteq \tau$. By induction hypothesis, $\Gamma \vDash_l l'{:}\tau_1$. Therefore by the rule $(l.y^t)$, $\Gamma \vDash_l l'.y{:}\tau$ as required.

- The case $l = *e$: in this case $l' = *e'$. By the rule $(*e^s)$, it is true that $e \rightsquigarrow^e_{(m,i)} e'$. Now since $\Gamma \vDash_l *e{:}\tau$, then by rule $(*e^t)$ there exists $\tau_1$ such that $\Gamma \vDash_e e{:}\tau_1$ and $\tau_1 = ptr^m \tau$. By Lemma 3, $\Gamma \vDash_e e'{:}ptr^m \tau$. Therefore by the rule $(*e^t)$, $\Gamma \vDash_l *e'.y{:}\tau$ as required.

**Lemma 3.** *Suppose that* $e \rightsquigarrow^e_{(m,i)} e'$ *and* $\Gamma \vDash_e e{:}\tau$. *Then* $\Gamma \vDash_e e'{:}\tau$.

**Proof.** The proof is by structure induction on the structure of expressions, $e$.

Some cases are shown below.

- The case $e = l$: in this case, by the rule $(l^s)$, $l \rightsquigarrow^l_{(m,i)} l'$ and $e' = l'$. By Lemma 2, $\Gamma \vDash_l l{:}\tau$ implies $\Gamma \vDash_l l'{:}\tau$. Hence by the rule $(l^t)$, $\Gamma \vDash_e e'{:}\tau$ as required.

- The case $e = e_1 \; i_{op} \; e_2$: in this case, by the rule $((e_1 \; i_{op} \; e_2)^t)$, $\tau = int(r_i, M)$. Moreover $\Gamma \vDash_e e_1{:}int(r_i, M)$ and $\Gamma \vDash_e e_2{:}int(r_i, M)$. Also by the rule $((e_1 \; i_{op} \; e_2)^s)$, we have $e_1 \rightsquigarrow^e_{(m,i)} e'_1$ and $e_2 \rightsquigarrow^e_{(m,i)} e'_2$. Hence by induction hypothesis $\Gamma \vDash_e e'_1{:}int(r_i, M)$ and $\Gamma \vDash_e e'_2{:}int(r_i, M)$. Therefore by the rule $((e_1 \; i_{op} \; e_2)^t)$, $\Gamma \vDash_e e'_1 \; i_{op} \; e'_2{:}int(r_i, M) = \tau$, as required.

- The case $e = modify-w(e,d)$: in this case, by the rule $(modify-w^s)$, $e' = modify-w(e',d)$ and $e \rightsquigarrow^e_{(m,i)} e'$. By the rule $(modify-w^t)$, it is true that $\Gamma \vDash_e e{:}ptr^m \tau$. Hence by induction hypothesis $\Gamma \vDash_e e'{:}ptr^{m'} \tau$. Therefore $\Gamma \vDash_e modify-w(e',d){:}\tau$, by the rule $(modify-w^t)$ as required for this case.

- The case $e = cast < int(r_j, M_j) \to int(r_i, M_i) > e$: in this case, by the rule $(cast^s_1)$, $e' = cast < int(r_j, M_j) \to int(r_i, M_i) > e'$ and $e \rightsquigarrow^e_{(m,i)} e'$. By the rule $(cast^t_1)$, it is true that $\Gamma \vDash_e e{:}int(r_j, M_j)$. Hence by induction hypothesis $\Gamma \vDash_e e' : int(r_j, M_j)$. Therefore $\Gamma \vDash_e cast < int(r_j, M_j) \to int(r_i, M_i) > e' : int(r_j, M_j)$ , by the rule $(cast_1^t)$. This completes the proof for this case.

Corollary 2 results directly from Lemma 3.

**Corollary 2.** *Suppose that* $e \rightsquigarrow_{(m,i)}^e e'$. *Then* $width - f(e) = width - f(e')$.

$S = compute\ S'\ at\ n$ and $S \rightsquigarrow_{(m,i)}^s S'$. Also since $\Gamma \vDash_s compute\ S\ at$ n: WT, it is true that $\Gamma \vDash_s S$ : WT, by the rule $(compute^t)$. Now by induction hypothesis, we conclude $\Gamma \vDash_s S'$ : WT . Hence by $(compute^t)$, $\Gamma \vDash_s compute\ S'\ at$ n: WT . This completes the proof for this case.

- The case $S = S_1; S_2$: in this case, by the rule $(seq^s)$, $S' = S_1'; S_2'$, $S_1 \rightsquigarrow_{(m,i)} S_1'$, and $S_2 \rightsquigarrow_{(m,i)} S_2'$. Also since $\Gamma \vDash_s S_1; S_2$: WT, it

$$\overline{x \rightsquigarrow_{(m,i)}^l x.(r_i,m)}\ (x^s) \qquad \frac{l \rightsquigarrow_{(m,i)}^l l'}{l.y \rightsquigarrow_{(m,i)}^l l'.y}\ (l^s) \qquad \frac{e \rightsquigarrow_{(m,i)}^e e'}{* e \rightsquigarrow_{(m,i)}^l * e'}\ (* e^s)$$

$$\overline{skip \rightsquigarrow_{(m,i)}^s skip}\ (x^s) \qquad \frac{l \rightsquigarrow_{(m,i)}^l l' \qquad e \rightsquigarrow_{(m,i)}^e e'}{l := e \rightsquigarrow_{(m,i)}^s l' := e'}\ (:=^s)$$

$$\frac{S \rightsquigarrow_{(m,i)}^s S'}{compute\ S\ at\ n \rightsquigarrow_{(m,i)}^s compute\ S'\ at\ n}\ (compute^s)$$

$$\frac{S_1 \rightsquigarrow_{(m,i)}^s S_1' \qquad S_2 \rightsquigarrow_{(m,i)}^s S_2'}{S_1; S_2 \rightsquigarrow_{(m,i)}^s S_1'; S_2'}\ (seq^s)$$

$$\frac{e \rightsquigarrow_{(m,i)}^e e' \qquad S_t \rightsquigarrow_{(m,i)}^s S_t' \qquad S_f \rightsquigarrow_{(m,i)}^s S_f'}{if\ e\ then\ S_t\ else\ S_f \rightsquigarrow_{(m,i)}^s if\ e'\ then\ S_t'\ else\ S_f'}\ (if^s)$$

$$\frac{e \rightsquigarrow_{(m,i)}^e e' \qquad S_t \rightsquigarrow_{(m,i)}^s S_t'}{while\ e\ do\ S_t \rightsquigarrow_{(m,i)}^s while\ e'\ do\ S_t'}\ (while^s)$$

$$\frac{d \rightsquigarrow_{(m,i)}^e d' \qquad S \rightsquigarrow_{(m,i)}^s S'}{dS \rightsquigarrow_{(m,i)}^p d'S'}\ (prog^m)$$

**Fig. 8.** Left expression, statement, and program slicing over machines.

**Theorem 1.** *Suppose that* $S \rightsquigarrow_{(m,i)}^s S'$ *and* $\Gamma \vDash_s$ S: WT. *Then* $\Gamma \vDash_s S'$: WT.

**Proof.** The proof is by structure induction on structure of statements, $S$. some cases are shown below.

- The case $S = l := e$: in this case, by the rule $(:=^s)$ , $S' = l' := e'$ . Moreover $l \rightsquigarrow_{(m,i)}^l l'$ and $e \rightsquigarrow_{(m,i)}^e e'$. Also since $\Gamma \vDash_s$ l := e: WT, it is true that $\Gamma \vDash_e$ e: τ and $\Gamma \vDash_l$ l: τ for some τ, by the rule $(:=^t)$. Now by Lemmas 2 and 3, we conclude $\Gamma \vDash_e e'$: τ and $\Gamma \vDash_l l'$: τ . Hence by $(:=^t)$, $\Gamma \vDash_s l' := e'$: WT, as required.
- The case $S = compute\ S\ at\ n$: in this case, by the rule $(compute^s)$ ,

is true that $\Gamma \vDash_s S_1$: WT and $\Gamma \vDash_s S_2$: WT, by the rule $(seq^t)$. Now by induction hypothesis, we conclude $\Gamma \vDash_s S_1'$: WT and $\Gamma \vDash_s S_2'$: WT. Hence by $(seq^t)$, $\Gamma \vDash_s S_1'; S_2'$: WT which completes the proof for this case.

Using Theorem 1 and Corollary 1, it is straightforward to conclude Corollary 3.

**Corollary 3.** *(Soundness of program slicing) Suppose that* $dS \rightsquigarrow_{(m,i)}^p d'S'$ *and* $\Gamma \vDash_p$ dS: WT. *Then* $\Gamma \vDash_p d'S'$: WT.

**Remark 3.** The type system of Section 2 can be realized as static semantics of the language $\mathcal{DLang}$. The proofs of Lemma 2 and 3 appear to rely on each other. This is absolutely true as expressions of a program are finite. The source of this sort of recursion is the syntactic structures of expressions and left expressions (Figure 2).

## 4. Related Work

Program slicing [23, 3] is a technique that enables a method to focus on certain part of a program. At a specific program point and with reference to a group of certain variables, a slice is an executable collection of program statements that maintain the original program behavior. Program slicing has many applications like parallelization [19], debugging [22], program comprehension [14], testing [16], downsizing, and restructuring. Statements deletions are the bases of the original concept [21] of program slice. However, there are many variants of this notion such as quasi static slicing [1, 17], dynamic slicing [22], conditioned slicing [18], and simultaneous dynamic slicing [7]. Other concepts [8] of slicing are based on generic notions of transformation such as simple statement deletion.

Typically, a slice is built on slicing criterion concept which is a pair $< p, V >$ of a program point and $V$ is a collection of variables. Hence at a program point $p$ and with reference to $V$, a slice that is based on $< p, V >$ is an executable collection of program statements that maintain the original program behavior. The maintainability here means that values of variables of $V$ are the same for the slice and the original program at the program point $p$. The concept of static slicing referees to maintaining the behavior of the original program on any input. However other forms of slicing maintain the behavior for a subset of program inputs.

Quasi static slicing [1, 17] is a hybrid technique for slicing that associates static and dynamic slicing [7]. Such hybrid techniques are required when analyzing programs that have fixed input variables and varying input values. Therefore on a set of potential program inputs, a quasi slice keeps the program behavior w.r.t. slicing variables. Potential value combinations assumed by unconstrained input variables specify the set of potential program inputs. Interestingly, the quasi static slice amounts to a static slice when all variables are unconstrained.

An alternative slicing approach is dynamic slicing [12, 11]. In this technique a dynamic analysis is used to find statements affected by a certain set of variables, on a specific anomalous execution path. This approach results in a considerable reduction in the size of the slice, and hence facilities bugs allocation. Moreover, dynamic slicing treats pointer variables and arrays in a practical way (in terms of run-time). Rather than treating every use (definition) of an array element as a use (definition) of the full array [2], dynamic slicing separately treats every array element. Equivalently, all along the execution of a program, dynamic slicing recognizes objects referenced by pointer variables. Interestingly, the quasi static slice amounts to a dynamic slice if all variable inputs are fixed.

A general version of slicing based on statement deletion is conditioned slicing [18]. This approach uses a slicing pattern for a collection of program executions to represent the original program behavior using only a collection of program statements. On the input, the first order logic is hence used to describe initial states of the program that specify these program executions.

Simultaneous dynamic program slicing [7, 23] constructs slices with reference to a collection of program executions. This approach is an extension of dynamic slicing in the form of a simultaneous application of dynamic slicing to a group of test cases, instead of only one case. However, on a group of test cases, a simultaneous program slice does not amount to applying dynamic slicing on the concerned test cases. Moreover, this multi-application of dynamic slicing is unsound in the sense that the simultaneous validity is not maintained on all the inputs. Simultaneous dynamic slicing is typically achieved iteratively, beginning with a group of statements. Hence simultaneous dynamic slices are built incrementally, via computations in each iteration.

## Acknowledgements:

Foundation item: Al Imam University (IMSIU) Project (No.: 330917). Author is grateful to Al Imam University (IMSIU), KSA for financial support to carry out this work.

4/3/2013